\documentclass[aps,prl,superscriptaddress]{revtex4-2}

\usepackage{amsmath}
\usepackage{graphicx}
\usepackage{MnSymbol}
\usepackage{parskip}
\usepackage{makecell}
\usepackage{booktabs}
\usepackage{multirow}
\usepackage{xcolor}

\begin{document}

\title{On the flash temperature in accelerated sliding contacts}

\author{B.N.J. Persson}
\affiliation{Peter Gr\"unberg Institute (PGI-1), Forschungszentrum J\"ulich, 52425, J\"ulich, Germany}
\affiliation{State Key Laboratory of Solid Lubrication, Lanzhou Institute of Chemical Physics, Chinese Academy of Sciences, 730000 Lanzhou, China}
\affiliation{MultiscaleConsulting, Wolfshovener str. 2, 52428 J\"ulich, Germany}

\begin{abstract}
The temperature increase in the contact regions between solids in sliding contact
can easily reach several hundred Kelvin and thereby dramatically affect friction and wear. 
Here I extend an earlier multiscale theory for the flash temperature (Ref. \cite{MP}) to the case
of accelerated motion, and present numerical results illustrating the theory.
\end{abstract}

\maketitle

\setcounter{page}{1}
\pagenumbering{arabic}

%\pagestyle{empty}

%%%%%%%%%%%%%% main text %%%%%%%%%%%%%%%%
%\begin{multicols}{2}

%%%%%%%%%%%%%% main text %%%%%%%%%%%%%%%%

{\bf 1 Introduction}

Friction between surfaces generates heat, leading to temperature increases at the contact points.
This phenomenon is known as flash temperature, which is the high, localized, and brief temperature spike that
occurs at the true points of contact between two rubbing solids. The rapid generation of heat at these locations
causes thermal spikes, resulting in intense flash temperatures as kinetic energy is converted into
heat. These spikes can be extremely high, sometimes reaching
over $1000\,^\circ {\rm C}$, but they are also incredibly brief,
lasting only for the instant that the asperities are in contact.
This process is so rapid that the generated heat has little time to conduct away into the bulk of the materials,
trapping thermal energy and further elevating the temperature at the contact points.

In almost all cases, most of the dissipated energy in sliding friction ends up as thermal energy within the sliding 
solids. The temperature field in the solids can be written as $T({\bf x},t) = T_0({\bf x},t) + \Delta T({\bf x},t)$.
The {\it background} temperature $T_0({\bf x},t)$ varies slowly in space and time while the {\it flash} temperature
$\Delta T({\bf x},t)$, varies rapidly in space and time. 
$\Delta T({\bf x},t)$ is non-zero only close to the asperity contact regions and is highly localized in space.

Frictional heating is important in a wide range of applications, spanning from ice and rubber friction to the
sliding of minerals like granite, which is central to earthquake dynamics.  The flash temperature
occurs in the asperity contact regions, where frictional energy is converted into heat. This localized heating
can have a crucial influence on the resulting friction, usually reducing it. This is the case for sliding ice,
where the flash temperature can melt the ice \cite{ice1,ice2,ice3}, or at least shift the temperature in the contact region towards the
melting point \cite{ice4,ice5}, thereby reducing the frictional shear stress and the friction force.
The same is true for granite
sliding on granite, where the melting point of the mineral (primarily quartz) may be reached at the sliding speeds (on the order of $\sim 1 \ {\rm m/s}$)
involved in earthquakes \cite{earth1,earth2,earth3}. Rubber friction depends exponentially on the temperature, and an increase in temperature
shifts the friction coefficient master curve to higher sliding speeds, which usually reduces the friction but sometimes
increases it \cite{rub1,rub2,rub3}.

\begin{figure}
\includegraphics[width=0.27\textwidth,angle=0.0]{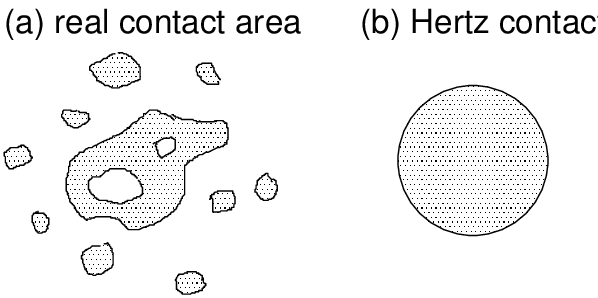}
\caption{\label{MacroAsperityTopSide.eps}
(a) A macroasperity contact area (schematic), and (b) the contact area in Hertz approximation.
}
\end{figure}

In Ref. \cite{MP} we have present a multiscale theory for the flash temperature during steady sliding.
The theory was based on the study of  temperature-temperature and temperature-stress correlation functions and included all relevant length scales.
In the limiting case of roughness on a single length scale, the theory reduces to the classical
theories of Jaeger, Archard, and Greenwood \cite{flash1,flash2,flash3} (see also \cite{flash4,flash5,flash6,TianGreen,ReddyhoffHardening,ZhuNumerical}).
These authors studied the temperature resulting from moving heat sources with circular shape with constant or Hertz-like pressure profiles. 
However, in reality the asperity contact regions have complex shapes where the macroasperity contact regions
consist of agglomerates of much smaller contact regions (see Fig. \ref{MacroAsperityTopSide.eps}) with complex
internal pressure distributions. In Ref. \cite{MP} it was shown
that for real surfaces with multiscale roughness the classical theories for the flash temperature fail severely.

Here I will extend the theory presented in Ref. \cite{MP} to accelerated motion.
In what follows all temperatures refer to the {\it increase} in the temperature above the background temperature $T_0(t)$.
Thus $T_{\rm flash}(t)$ is a weighted average flash temperature,
and the actual temperature in the contact regions is $T_0(t)+T_{\rm flash}(t)$. Similarly,
the temperature-stress correlation function $\langle T({\bf x},t)\sigma ({\bf x},t)\rangle$ is calculated with the background temperature
$T_0$ subtracted from $T({\bf x})$. Stated differently, all temperatures refer to actual temperatures only if the background
temperature vanishes.

\vskip 0.2cm
{\bf 2 Theory}

The temperature distribution in a sliding contact is considered for accelerated motion.
We assume that frictional energy is produced only in the 
real contact area within a nanometer-thin surface layer, which we consider to be infinitesimally thin.
We analyze the heat diffusion in the half-space $z>0$ with a heat source (energy per unit area and unit time)
$\dot q ({\bf x},t) = \dot q (x,y,t)$ on the surface $z=0$, and with the heat current $J_z$ vanishing
on the regions of the surface where $\dot q({\bf x},t)=0$. This
heat diffusion problem can be solved most easily by extending the problem to the heat diffusion
in an infinite solid with a heat source on the surface $z=0$, which is twice as strong, $2 \dot q ({\bf x},t)$.
Thus, the temperature distribution in the sliding solid is determined by the heat diffusion equation
$$\rho c_{\rm P} {\partial T \over \partial t} -  \kappa \nabla^2 T  = 2 \dot q ({\bf x},t) \delta (z),$$
where $\rho$, $c_{\rm P}$, and $\kappa$ are the mass density, the specific heat capacity, and the thermal conductivity,
respectively. Introducing the thermal diffusivity $D=\kappa/\rho c_{\rm P}$, we can write
$${\partial T \over \partial t} -  D \nabla^2 T  = {2 D \over \kappa } \dot q ({\bf x},t) \delta (z), \eqno(1)$$
where $\dot q $ depends on the surface position ${\bf x}=(x,y)$ and time $t$. 

The heat source $\dot q$ is non-vanishing only in the area of real contact, which is the area where the normal stress is non-zero.
In the simplest case, one can assume that $\dot q$ is proportional to the normal stress $\sigma ({\bf x})$ so that
$$\dot q ({\bf x},t) = \dot q({\bf x}-{\bf r} (t))= \mu(t) v(t) \sigma ({\bf x} - {\bf r}(t)) = Q(t) \sigma ({\bf x} - {\bf r}(t))$$
where we define $Q(t) = \mu (t) v(t)$ and where ${\bf r}(t)$ is the center of mass position of the sliding surface. 

Writing
$$\sigma ({\bf x})  = \int d^2 q \, \sigma ({\bf q}) e^{i {\bf q} \cdot {\bf x}}$$
we get
$$\dot q  ({\bf x},t) = Q(t) \int d^2 q \, \sigma({\bf q}) e^{i {\bf q} \cdot [{\bf x}-{\bf r}(t)]} =  Q(t) \int d^2 q \, \sigma({\bf q},t) e^{i {\bf q}\cdot {\bf x}} \eqno(2)$$
where
$$\sigma({\bf q},t) =\sigma({\bf q}) e^{-i {\bf q} \cdot {\bf r}(t)}\eqno(3)$$
Using the identity
$$\delta (z) = {1\over 2 \pi} \int dk \, e^{ikz} $$
we get
$$\dot q({\bf x},t) \delta(z) = {Q(t) \over 2 \pi} \int d^2 q dk \, \sigma({\bf q}) e^{i {\bf q} \cdot [{\bf x}-{\bf r}(t)] +ikz}\eqno(4)$$
Writing
$$T({\bf x},z,t) = \int d^2 q dk \, T({\bf q},k,t) e^{i ({\bf q} \cdot {\bf x} +kz)}\eqno(5)$$
from (1) we get after Fourier transformation of the $({\bf x},z)$ dependency:
$${\partial T \over \partial t} +D(q^2+k^2) T = {D \over \pi \kappa} Q(t) \sigma({\bf q}) e^{-i {\bf q} \cdot {\bf r}(t)}$$
or
$${\partial \over \partial t} \left (e^{D(q^2+k^2)t} T \right ) = {D \over \pi \kappa} Q(t) \sigma({\bf q}) e^{D(q^2+k^2)t} e^{-i {\bf q} \cdot {\bf r}(t)}$$
or
$$T({\bf q},k,t) = {D \over \pi \kappa} \int_0^t dt' \, Q(t') \sigma({\bf q}) e^{-D(q^2+k^2)(t-t')}  e^{-i {\bf q} \cdot {\bf r}(t')}\eqno(6)$$
where we have assumed the sliding motion start at $t=0$ so that the flash temperature $T=0$ at $t=0$.
Using that
$$\int_{-\infty}^\infty dk \, e^{-Dk^2 (t-t')} e^{ikz} =\left [  {\pi \over D(t-t') } \right ]^{1/2} e^{-z^2/[4D(t-t')]}$$
we get
$$T({\bf q},z,t) = { \surd D \over \surd \pi \kappa}  \int_0^t dt' \, {Q(t') \over (t-t')^{1/2}} \sigma({\bf q}) e^{-Dq^2(t-t')}  e^{-i {\bf q} \cdot {\bf r}(t')-z^2/[4D(t-t')]}\eqno(7)$$
Here we are interested in the surface temperature ($z=0$) which we denote by $T({\bf x},t)$ with the Fourier transform: 
$$T({\bf q},t) = { \surd D \over \surd \pi \kappa}  \int_0^t dt' \, {Q(t') \over (t-t')^{1/2}} \sigma({\bf q}) e^{-Dq^2(t-t')}  e^{-i {\bf q} \cdot {\bf r}(t')}\eqno(8)$$

In applications to sliding friction, the temperature within the asperity contact regions is of primary importance, 
rather than the temperature distribution across the entire surface. To isolate this, one can define an effective 
flash temperature by weighting the temperature by the local contact stress:
$$T_{\rm flash} (t) = \frac{\langle T({\bf x},t) \sigma ({\bf x},t) \rangle}{\langle \sigma ({\bf x},t) \rangle}, \eqno(9)$$
where $\langle \sigma ({\bf x},t) \rangle = \sigma_0$ is the nominal contact pressure. 
Defined this way, the temperature is weighted only where the stresses are high, and is ignored in the non-contact regions 
where $\sigma(\mathbf{x},t) = 0$. This is the optimum definition of the average or effective friction coefficient, which can be used
in friction calculations with a temperature depending friction coefficient $\mu = \mu(T)$ with $T=T_0+T_{\rm flash}$

Using the relation
$$\langle T({\bf x},t) \sigma ({\bf x},t) \rangle 
= \frac{(2\pi )^2}{A_0} \int d^2q \, \langle T({\bf q},t) \sigma (-{\bf q},t) \rangle \eqno(10)$$
and (3) and (8) we obtain:
$$\langle T({\bf x},t) \sigma ({\bf x},t) \rangle = { \surd D \over \surd \pi \kappa}  \int_0^t dt' \, {Q(t') \over (t-t')^{1/2}} 
\frac{(2\pi )^2}{A_0} \int d^2q \,  \langle \sigma({\bf q}) \sigma(-{\bf q}) \rangle e^{-Dq^2(t-t')}  
e^{i {\bf q} \cdot [{\bf r}(t)-{\bf r}(t')]}\eqno(11)$$
The stress power spectrum \cite{Persson1}
$$\langle \sigma ({\bf q}) \sigma (-{\bf q})\rangle = {A_0\over (4\pi)^2} (E^*)^2  q^2 C(q) W(q), \eqno(12)$$
where $E^* = E/(1-\nu^2)$ is the effective modulus,
where $C(q)$ is the surface roughness power spectrum, and
$$W(q) = P(q) [\gamma + (1-\gamma)P^2(q)],\eqno(13)$$
where
$$P(q)={\rm erf}\left ({\sigma_0 \over 2 \surd G}\right )\eqno(14)$$
$$G(q)= {\pi \over 4} ( E^*)^2 \int_{q_0}^q dq' \ {q'}^3 C(q'),\eqno(15)$$
Using (11) and (12) gives
$$\langle T({\bf x},t) \sigma ({\bf x},t) \rangle = { \surd D \over  4\surd \pi \kappa} (E^*)^2 \int_0^t dt' \, {Q(t') \over (t-t')^{1/2}}  \int d^2q \, q^2 C(q) W(q)  e^{-Dq^2(t-t')}  
e^{i {\bf q} \cdot [{\bf r}(t)-{\bf r}(t')]}\eqno(16)$$
If we assume sliding along the $x$-axis then ${\bf q} \cdot [{\bf r}(t)-{\bf r}(t')] = q [x(t)-x(t')] {\rm cos}\phi$
and (16) becomes
$$\langle T({\bf x},t) \sigma ({\bf x},t) \rangle = { \surd D \over  4\surd \pi \kappa} (E^*)^2 \int_0^t dt' \, {Q(t') \over (t-t')^{1/2}}  \int d^2q \, q^2 C(q) W(q)  e^{-Dq^2(t-t')}  
e^{i q[x(t)-x(t')] {\rm cos}\phi}\eqno(17)$$
Using that
$${1\over 2 \pi} \int_0^{2\pi} d\phi \, e^{i x {\rm cos}\phi} = J_0(x), $$
where $J_0(x)$ is the zero order Bessel function, (17) gives
$$T_{\rm flash} (t) =   {\surd (\pi D) \over 2 \kappa \sigma_0} (E^*)^2 \int_0^t dt' \, {v(t') \mu (t') \over (t-t')^{1/2}} 
  \int_{q_0}^{q_1} dq \, q^3 C(q) W(q)  e^{-Dq^2(t-t')} J_0(q[x(t)-x(t')]) \eqno(18)$$
%Here we indicate that the nominal contact pressure $\sigma_0(t)$ may depend on time, which also introduce a time dependency of
%$P(q)$ and hence of $W(q)$, which we now denote as $W(q,t)$.
For a constant sliding speed $x=vt$ after long enough sliding distance (after run-in; steady-state) where
$\mu(t)$ is time independent, equation (18) gives
$$T_{\rm flash} (t) =   {\surd (\pi D) \over 2 \kappa \sigma_0} (E^*)^2  v \mu   \int_0^t dt' \, {1 \over (t-t')^{1/2}} 
  \int_{q_0}^{q_1} dq \, q^3 C(q) W(q) e^{-Dq^2(t-t')} J_0(qv[t-t']) \eqno(19)$$
This time dependent part of this expression is
$$K = \int_0^t dt' \, {1 \over (t-t')^{1/2}} e^{-Dq^2(t-t')} J_0(qv[t-t'])$$
$$= {1\over 2 \pi} \int_0^{2\pi} d\phi \int_0^t dt' \, {1 \over (t-t')^{1/2}} e^{-Dq^2(t-t')} e^{iqv[t-t']{\rm cos}\phi }$$
Writing $t-t'= w$ we get
$$K={1\over 2 \pi} \int_0^{2\pi} d\phi \int_0^t dw {1 \over w^{1/2}} e^{-[Dq^2 -iqv {\rm cos}\phi] w }$$
Writing $w=y^2$ and taking the (steady-state) limit $t\rightarrow \infty$ we get
$$K = {1\over \pi} \int_0^{2\pi} d\phi \int_0^\infty dy \,  e^{-[Dq^2-iqv \, {\rm cos}\phi]y^2}$$
Using the standard integral
$$\int_0^\infty dy \, e^{-a y^2} = {1\over 2} \left ({\pi \over a}\right )^{1/2}$$
we get
$$K= {1\over 2 \surd \pi} \int_0^{2\pi} d\phi  \, {1\over \left (Dq^2 - iqv \, {\rm cos}\phi \right )^{1/2}}$$
and (19) becomes
$$T_{\rm flash} =   {\surd D \over 4 \kappa \sigma_0(t)} (E^*)^2  v \mu  
  \int_{q_0}^{q_1} dq \, q^3 C(q) W(q) \int_0^{2\pi} d\phi  \, {1\over \left (Dq^2 - iqv \, {\rm cos}\phi \right )^{1/2}}$$
or
$$T_{\rm flash}  =   {v \mu \over \kappa \sigma_0} (E^*)^2   
  \int_{q_0}^{q_1} dq \, q^2 C(q) W(q) {\rm Re} \int_0^{\pi/2} d\phi  \, {1\over \left (1 - i(v/Dq) \, {\rm cos}\phi \right )^{1/2}}\eqno(20)$$
which agree with the result derived in Ref. \cite{MP} for steady sliding.

Following Ref. \cite{MP} we define 
$$\nabla T_{\rm flash} (t) = {\langle \nabla T({\bf x},t) \sigma ({\bf x},t) \rangle \over \langle \sigma ({\bf x},t) \rangle}, \eqno(21)$$
and
$$\nabla^2 T_{\rm flash} (t) = {\langle \nabla^2 T({\bf x},t) \sigma ({\bf x},t) \rangle \over \langle \sigma ({\bf x},t) \rangle}, \eqno(22)$$
By choosing the $x$-axis along the sliding direction, only the $x$-component of (20) is non-vanishing. Denoting this with $T'_{\rm flash}$
(where the prime indicate derivative with respect to $x$), we find
$$\langle \partial_x T({\bf x},t) \sigma ({\bf x},t) \rangle = 
{ \surd D \over  4\surd \pi \kappa} (E^*)^2 \int_0^t dt' \, {Q(t') \over (t-t')^{1/2}}  \int d^2q \, q^2 C(q) W(q)  e^{-Dq^2(t-t')}  
[iq {\rm cos} \phi ] e^{i q[x(t)-x(t')] {\rm cos}\phi}$$
Using that
$${1\over 2 \pi} \int_0^{2\pi} d\phi \, {\rm cos}\phi e^{i x {\rm cos}\phi} = {1\over 2 \pi i} {d \over dx} \int_0^{2\pi} d\phi \, e^{i x {\rm cos}\phi} = -i J_0'(x) =i J_1(x) $$
we get
$$T'_{\rm flash} (t) =   - {\surd (\pi D) \over 2 \kappa \sigma_0} (E^*)^2 \int_0^t dt' \, {v(t') \mu (t') \over (t-t')^{1/2}} 
  \int_{q_0}^{q_1} dq \, q^4 C(q) W(q)  e^{-Dq^2(t-t')} J_1(q[x(t)-x(t')]) \eqno(23)$$
Similarly
$$\nabla^2 T_{\rm flash} (t) =   - {\surd (\pi D) \over 2 \kappa \sigma_0} (E^*)^2 \int_0^t dt' \, {v(t') \mu (t') \over (t-t')^{1/2}} 
  \int_{q_0}^{q_1} dq \, q^5 C(q) W(q)  e^{-Dq^2(t-t')} J_0(q[x(t)-x(t')]) \eqno(24)$$

Following Ref. \cite{MP} we define the width $D_{\rm flash}$ and the slope-length parameter $d_{\rm flash}$ by
$$D_{\rm flash} = \left [{-T_{\rm flash} \over \nabla^2 T_{\rm flash} }\right ]^{1/2}\eqno(25)$$
$$ d_{\rm flash} = {-T_{\rm flash} \over T'_{\rm flash}}\eqno(26)$$
For a Hertz contact region with radius $R$, $D_{\rm flash} = \alpha R$ and $ d_{\rm flash} =\beta R$. 
In Ref. \cite{FlashJCP} I have shown that for steady sliding at high sliding speed 
for the Hertz contact $\alpha \approx 0.45$ and $\beta \approx 1.36$, while for stationary contact
$\alpha \approx 0.63$ and $\beta = \infty$.  Here high sliding speed corresponds to $v \gg D/R$ where $D$ is the thermal diffusivity.

\begin{figure}
\includegraphics[width=0.47\textwidth,angle=0.0]{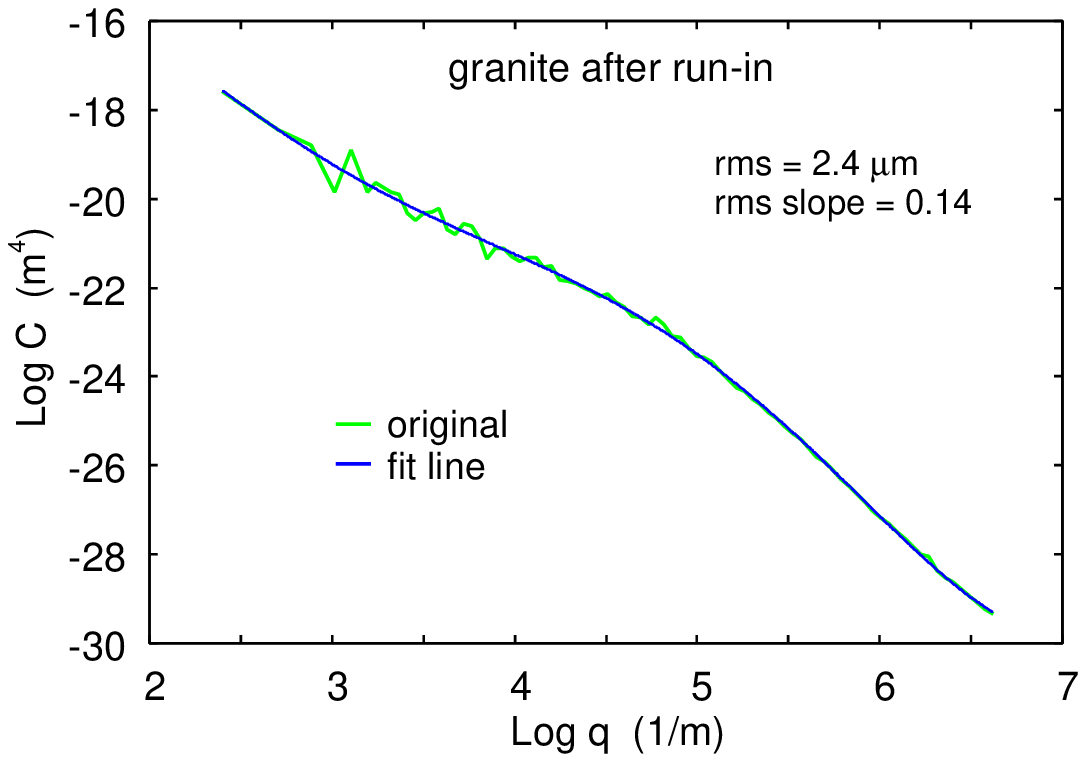}
\caption{\label{1logq.2logC.granite.after.runin.eps}
The surface roughness power spectrum of the granite surface used in the present study. The spectrum represents the run-in state and is shown as a function of the wavenumber $q$.
}
\end{figure}

\vskip 0.2cm
{\bf 3 Numerical results}

Here, we consider non-stationary sliding for quartz-on-quartz and silica-on-silica,
which are minerals of interest in earthquake dynamics, as the continental parts of tectonic
plates are often described as \textit{granitic} in overall composition, and granite consists mainly of quartz.
The same systems was studied in Ref. \cite{MP} but only for sliding at a constant speed (steady-state).
As in Ref. \cite {MP} we will use the measured surface roughness power spectrum of a run-in granite block and substrate
system shown in Fig. \ref{1logq.2logC.granite.after.runin.eps} (see Ref. \cite{earth3}).
After run-in, the contact regions deform mostly elastically at the length scales for which
the power spectrum has been measured (down to the micrometer length scale, which is the 
resolution of the stylus topography-instrument used). 
Thus, we will neglect plastic flow in the model calculations and will use the thermal parameters
for quartz and silica given in Table~\ref{thermal}.

%\begin{widetext}
%
\begin{table*}[hbt]
   \caption{The mass density $\rho$, the heat conductivity $\kappa_{\rm th}$, the heat capacity per unit mass $C_V$, and the heat diffusivity $D$
for silica and quartz.}
   \label{thermal}
   \renewcommand{\arraystretch}{1.5} % Default value: 1
   \begin{center}
      \begin{tabular}{@{}|l||c|c|c|c|@{}}
         \hline
            solid   &  $\rho \ {\rm (kg/m^3)}$  & $\kappa_{\rm th} \ {\rm (W/Km)}$ & $C_V \ {\rm (J/kgK)}$ & $D \  {\rm (m^2/s)}$ \\
         \hline
         \hline
            silica  & 2400 & 1.4 & 700 & $8.3 \times 10^{-7}$  \\
         \hline
            quartz  & 2400 & 11.7 & 700 & $7.0 \times 10^{-6}$  \\
         \hline
      \end{tabular}
   \end{center}
\end{table*}
%
%\end{widetext}

\begin{figure}
\includegraphics[width=0.47\textwidth,angle=0.0]{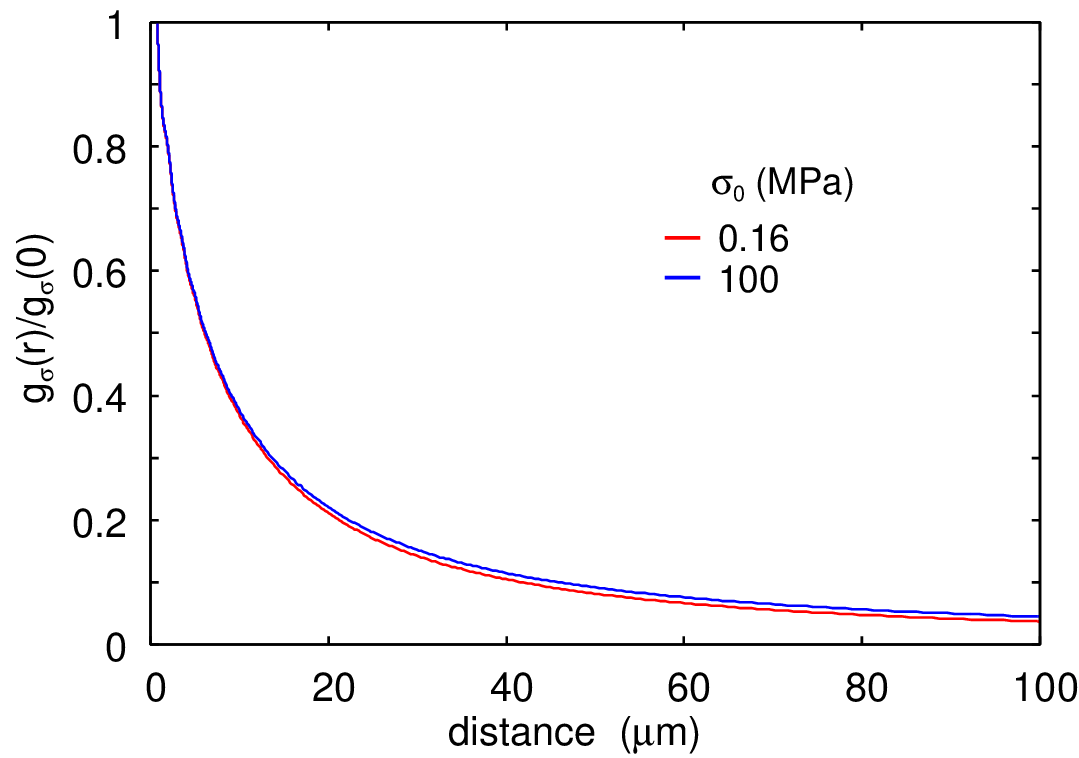}
\caption{\label{1distanceMicrometer.stresscorrelationNormalized.granite.eps}
The normalized stress correlation $g_{\sigma}(r)/g_{\sigma}(0)$ as a function of the separation $r=|{\bf x}-{\bf x}'|$
for the contact between two granite surfaces, using the elastic properties of quartz (which are nearly the same as those for silica).
The red and blue lines were obtained assuming nominal contact pressures of $0.16$ and $100$ MPa, respectively.
If the contact diameter $r_0$ were defined by $g_{\sigma}(r_0) = \alpha g_{\sigma}(0)$
with $\alpha = 0.2$, then we would get $r_0 \approx 20 \ {\rm \mu m}$. However, the macroasperity contacts
consist of a central region surrounded by
disconnected islands (see Ref. \cite{Miy}); these islands of contact
are elastically correlated with the central part of the stress correlation function
$g_{\sigma}(r)$, which results in a long tail of $g_{\sigma}(r)$ extending to larger distances $r$.
Adapted from Ref. \cite{MP}.
}
\end{figure}

\begin{figure}
\includegraphics[width=0.48\textwidth,angle=0.0]{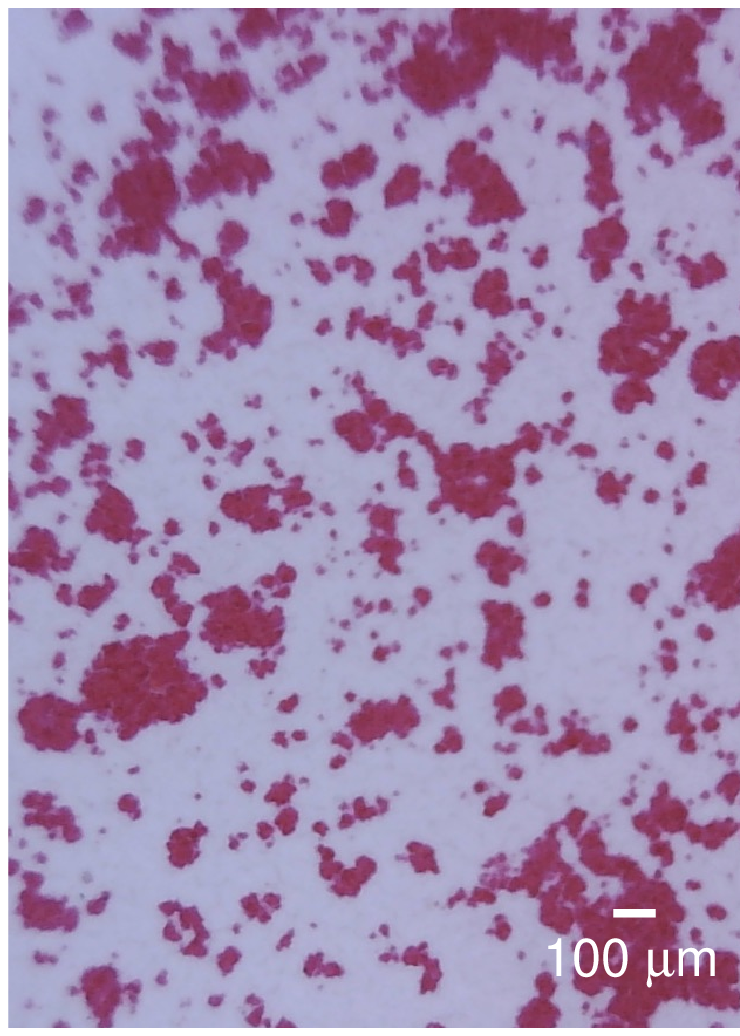}
\caption{\label{magni1.eps}
Picture of the contact regions between two granite surface obtained using a pressure-sensitive film
(Fujifilm, Super Low Pressure, $0.5-2.5 \ {\rm MPa}$ pressure range).
The nominal contact pressure is $\sigma_0 \approx 1 \ {\rm MPa}$.
It appears as if the contact regions occupy $\sim 10 \%$ of the nominal contact area
which would correspond to the average contact pressure $\sim 10 \ {\rm MPa}$.
However, contact mechanics calculations show that the contact pressure is close to the
yield stress of quartz which is several GPa (see Ref. \cite{earth3}). Hence the real contact area is less that
$1\%$ of what appears from the figure. The failure of the picture to show the true relative contact area 
is due to lateral resolution (pixel size) which is $\approx 30 \ {\rm \mu m}$,
which is the typical size of the red dots when a microcapsule with the ink is broken.
However, the figure shows that there are clusters of contact regions with diameter 
$ > 100 \ {\rm \mu m}$ which is consistent with the tail in the stress correlation function
which extend to separations $r> 100 \ {\rm \mu m}$.
%as indicated shematically in (b)
%where one small circular region in the apparent contact area is magnified.
%Upon magnification the contact area break-up into micrometer
%sized surface contact regions where the local pressure is close to the
%quartz plastic yield stress (of order GPa).
%The lateral resolution (pixel size) is $\lambda = 30 \ {\rm \mu m}$,
%which is the size of the red dot when a microcapsule with the ink is broken.
}
\end{figure}

Fig. \ref{1distanceMicrometer.stresscorrelationNormalized.granite.eps}
shows the calculated normalized stress correlation $g_{\sigma} (r)/g_{\sigma} (0)$ as a function of the separation $r=|{\bf x}-{\bf x}'|$
for the contact between two granite surfaces using the elastic properties of quartz (which are nearly the same as those for  silica).
If the effective contact diameter $d$ is defined by $g_{\sigma} (d) = \alpha g_{\sigma} (0)$
with $\alpha = 0.5$, then we would get $d \approx 6 \ {\rm \mu m}$. However, the macroasperity contacts
consist of a central compact region surrounded by
disconnected islands (see Ref. \cite{Miy}), and these islands of contact
are elastically correlated with the central part of the stress correlation function
$g_{\sigma} (r)$, which results in a long tail of $g_{\sigma} (r)$ extending to larger distances $r$.

Fig. \ref{magni1.eps} shows a picture of the contact area between two granite blocks obtained a pressure-sensitive film
(Fujifilm, Super Low Pressure, $0.5-2.5 \ {\rm MPa}$ pressure range).
The nominal contact pressure is $\sigma_0 \approx 1 \ {\rm MPa}$, and it appears as if the contact 
regions occupy $\sim 10 \%$ of the nominal contact area,
which would correspond to the average contact pressure $\sim 10 \ {\rm MPa}$.
However, contact mechanics calculations show that the contact pressure is close to the
yield stress of quartz which is several GPa (see Ref. \cite{earth3}). Hence the real contact area is less that
$1\%$ of what appears from the figure. The failure of the picture to show the true relative contact area 
is due to lateral resolution (pixel size) which is $\approx 30 \ {\rm \mu m}$,
which is the typical size of the red dots when a microcapsule with the ink is broken.
However, the figure shows that there are clusters of contact regions with diameter 
$ > 100 \ {\rm \mu m}$ which is consistent with the tail in the stress correlation function
(see Fig. \ref{1distanceMicrometer.stresscorrelationNormalized.granite.eps}) which extend to separations $r> 100 \ {\rm \mu m}$.

The red and blue lines in
Fig. \ref{1distanceMicrometer.stresscorrelationNormalized.granite.eps}
were obtained assuming nominal contact pressures of $0.16$ and $100 \ {\rm MPa}$, respectively.
Note that $g_{\sigma} (r)/g_{\sigma} (0)$ is nearly independent of the applied pressure, which holds true as long as the
relative contact area satisfies $A/A_0 \ll 1$. The reason is that increasing the nominal pressure $\sigma_0$ when $A/A_0 \ll 1$ increases
the number of macroasperity contact regions in proportion to $\sigma_0$, while the pressure distribution within the macroasperity contact
regions remains nearly unchanged. As long as $A/A_0 \ll 1$, the elastic coupling between the macroasperity contact regions is small,
and $g_{\sigma} (r)/g_{\sigma} (0)$ will nearly vanish for $r=|{\bf x}-{\bf x}'|$
larger than the diameter of the macroasperity contact regions. These statements require a large enough system
with a surface roughness power
spectrum that has a roll-off, which is always the case for surfaces of engineering interest.

\begin{figure}
\includegraphics[width=0.47\textwidth,angle=0.0]{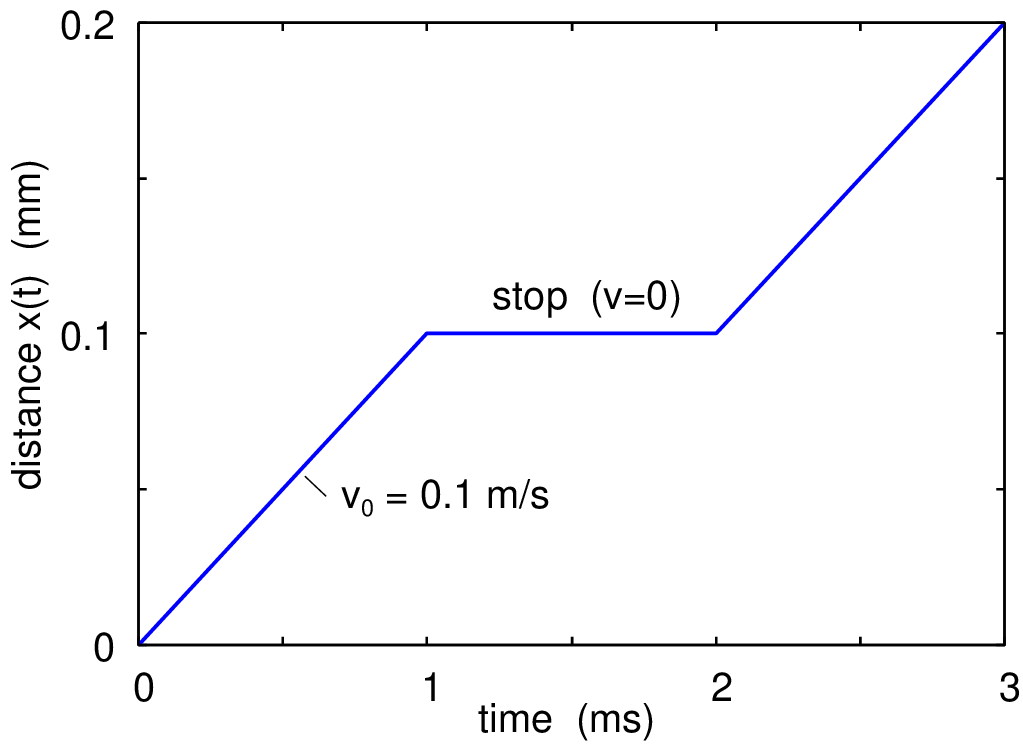}
\caption{\label{1time.2x.event.eps}
Sliding motion assumed in the flash temperature calculations.
The block start to slide at time $t=0$ with the speed $v_0 = 0.1 \ {\rm m/s}$.
After $1 \ {\rm ms}$ the motion stops and start again with the same velocity
after $2 \ {\rm ms}$.
}
\end{figure}

We will now present results for the flash temperature for the non-steady {\it sliding-stop-sliding} motion
shown in Fig. \ref{1time.2x.event.eps}. The block start to slide at time $t=0$ with the speed $v_0 = 0.1 \ {\rm m/s}$.
After $1 \ {\rm ms}$ the motion stops and start again at $t=2 \ {\rm ms}$ with the velocity $v_0 = 0.1 \ {\rm m/s}$.
The nominal contact pressure is $\sigma_0=100 \ {\rm MPa}$
and I assume the constant friction coefficient $\mu = 1.0$. In reality, for quartz
(or silica) the friction coefficient decreases continuously as the temperature approach the
quartz melting temperature \cite{earth1}, but this effect is not included here as it would make the
interpretation of the flash temperature results more complex.

\begin{figure}
\includegraphics[width=0.47\textwidth,angle=0.0]{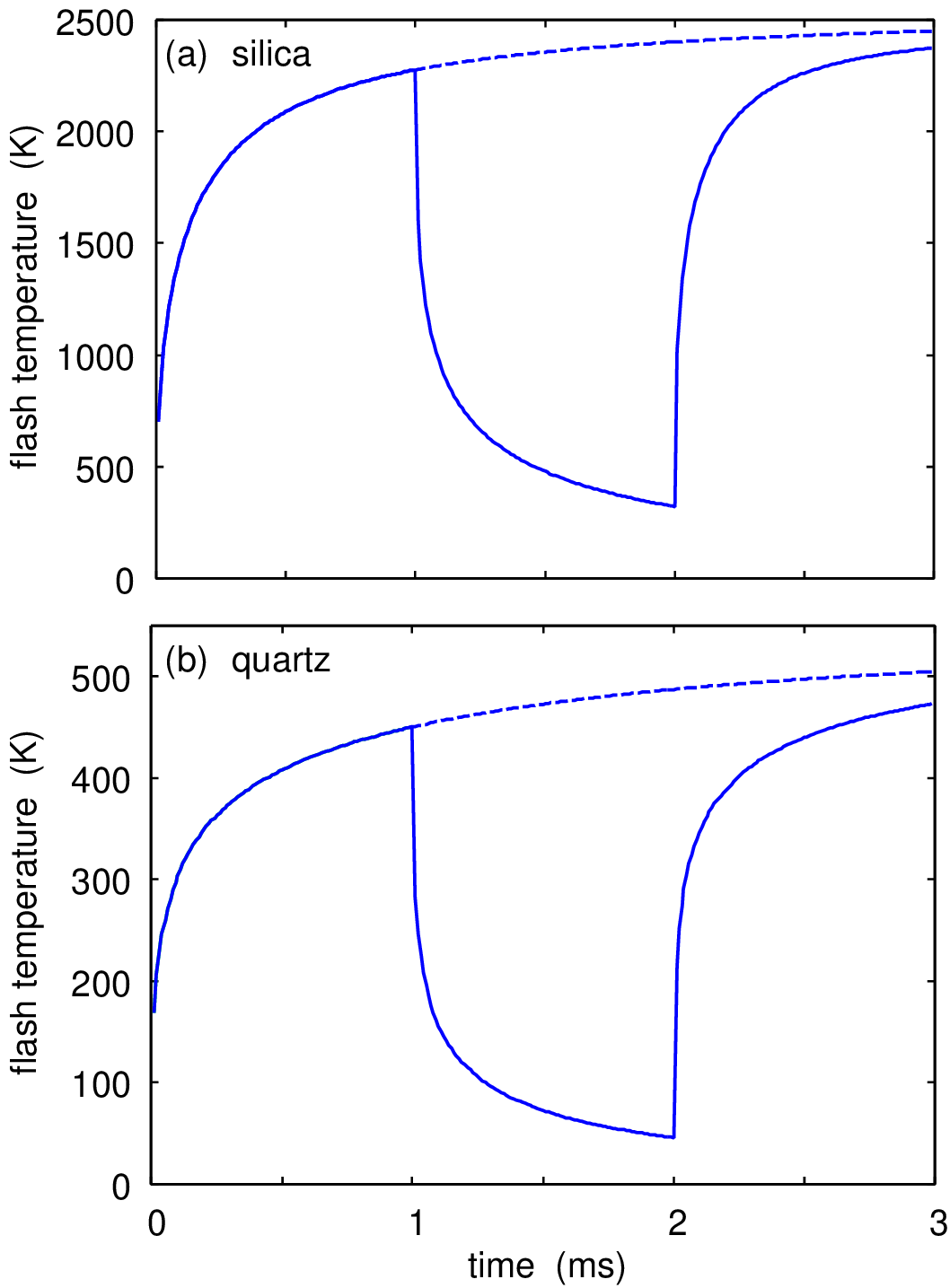}
\caption{\label{quartz.flash.sliding.eps}
The flash temperature $T_{\rm flash} (t)$ as a function of time for (a) silica
and (b) quartz. The dashed line is the flash temperature during continuous sliding
with the velocity $v(t)=v_0$ for $T>0$ while the solid line is for the 
sliding-stop-sliding event shown in Fig. \ref{1time.2x.event.eps} with $v_0 = 0.1 \ {\rm m/s}$.
}
\end{figure}

Fig. \ref{quartz.flash.sliding.eps} shows the flash temperature $T_{\rm flash} (t)$ as a function of time for (a) silica
and (b) quartz. The dashed lines are the flash temperature for sliding at the speed $v_0 = 0.1 \ {\rm m/s}$ for $t>0$.
For this case the flash temperature increases continuously during sliding and approach the steady state sliding value
for large time given by (19). The solid lines is for the {\it sliding-stop-sliding} event shown in Fig. \ref{1time.2x.event.eps}.
As expected, during {\it stop} the temperature decay monotonically with time. Note the fast increase in the
temperature at the onset of sliding both at the start of sliding and after the stop time period. Thus, the flash temperature
has reached $80\%$ of its steady state value (which is about $2500 \ {\rm K}$ for silica and $500 \ {\rm K}$ for quartz)
after $\Delta t \approx 0.5 \ {\rm ms}$. This correspond to a sliding distance of $v_0 \Delta t \approx 50 \ {\rm \mu m}$
and is consistent with the size of the macroasperity contact regions: the stress-stress correlation function for
$r=50 \ {\rm \mu m}$ is about $10\%$ of its initial value. That is, the flash temperature is close to its steady state value
after moving a distance of order a few times the average diameter of the contact regions deduced from the stress-stress
correlation function.

Fig. \ref{quartz.flash.sliding.eps} shows that the flash temperature is much higher for silica than for quartz.
This is due to the much higher thermal diffusivity for quartz compared to silica. This in turn is due to the disordered
lattice structure of silica, with reduces the phonon mean free path and hence the thermal conductivity.

\begin{figure}
\includegraphics[width=0.47\textwidth,angle=0.0]{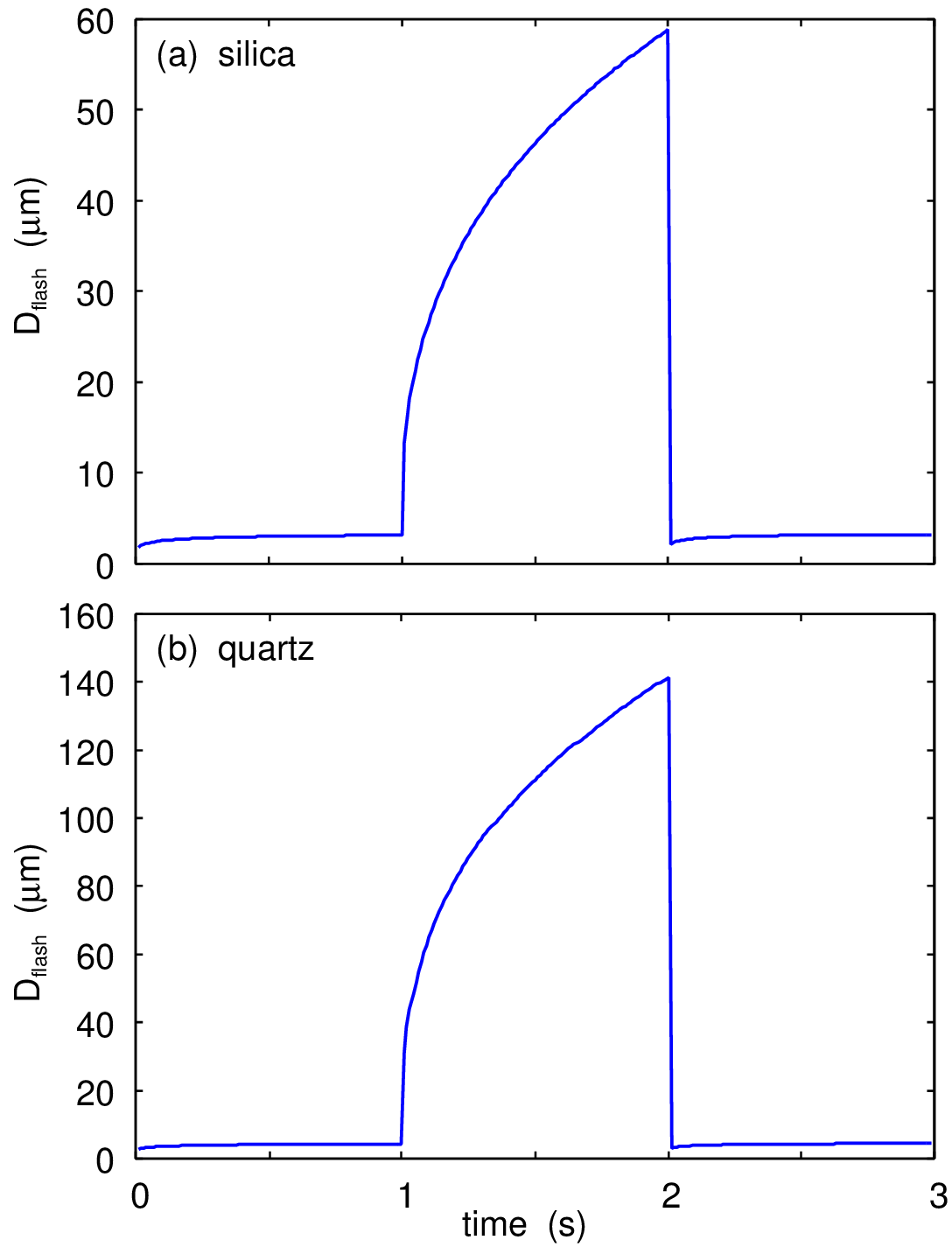}
\caption{\label{1time.2Dflash.quartz.eps}
The effective diameter of the hot spots $D_{\rm flash} (t)$ as a function of time for (a) silica
and (b) quartz.
}
\end{figure}

\begin{figure}
\includegraphics[width=0.47\textwidth,angle=0.0]{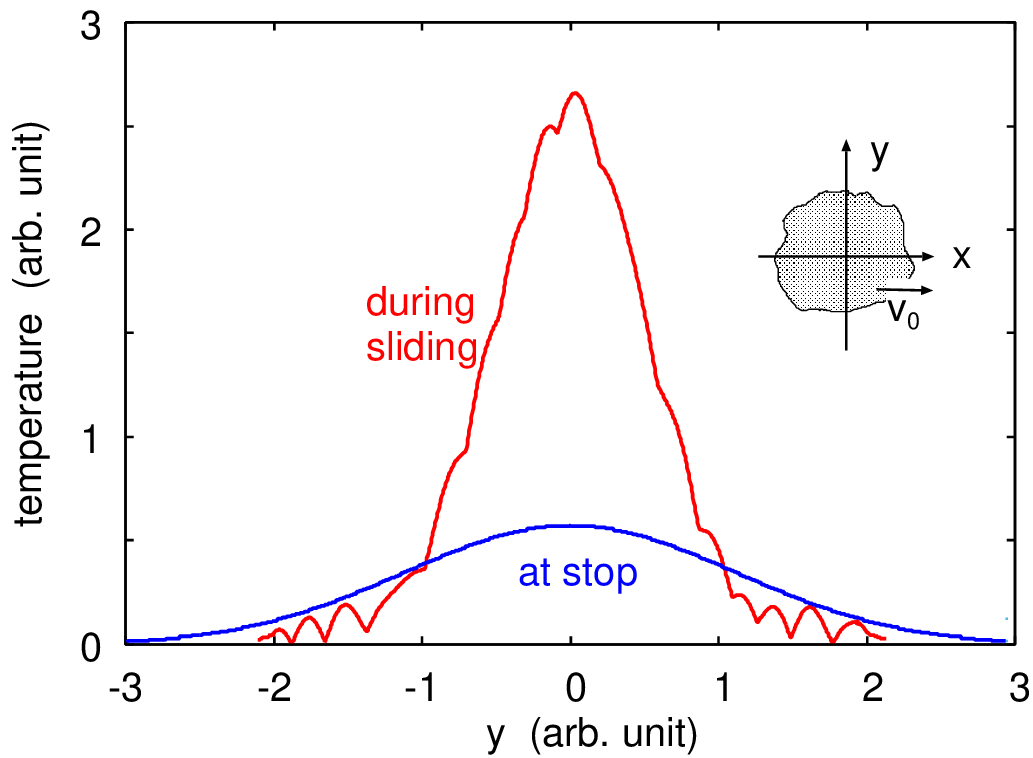}
\caption{\label{1x.2Gaus.eps}
The flash temperature distribution in the $y$-direction orthogonal to the sliding
direction during sliding (red) and after some stopping time (blue). Because of the thermal
heat diffusion the temperature distribution broaden during stop and the ``granularity'' in the
temperature distribution resulting from the microasperity contact regions get smoothed out
by the thermal diffusion. (Schematic.)
}
\end{figure}

Fig. \ref{1time.2Dflash.quartz.eps}
shows the effective diameter of the hot spots $D_{\rm flash} (t)$ as a function of time for (a) silica
and (b) quartz. The effective diameter during steady sliding is similar for silica and quartz
(about $3$ and $4 \ {\rm \mu m}$ for silica and quartz, respectively), and corresponds roughly to the
distance $r$ where the stress-stress correlation function has reached half of its value for $r=0$,
but during the stop time period $D_{\rm flash} (t)$ increases much faster with time for quartz than for silica
and reach a maximum which is more than $2$ times higher for quartz. This is again due to the higher thermal
diffusivity of quartz. 

Fig. \ref{1x.2Gaus.eps} illustrate schematically the 
flash temperature distribution in the $y$-direction orthogonal to the sliding
direction during sliding (red) and after some stopping time (blue). Because of the thermal
heat diffusion the temperature distribution broaden during stop and the ``granularity'' in the
temperature distribution resulting from the microasperity contact regions is smoothed 
by the thermal diffusion.

\begin{figure}
\includegraphics[width=0.47\textwidth,angle=0.0]{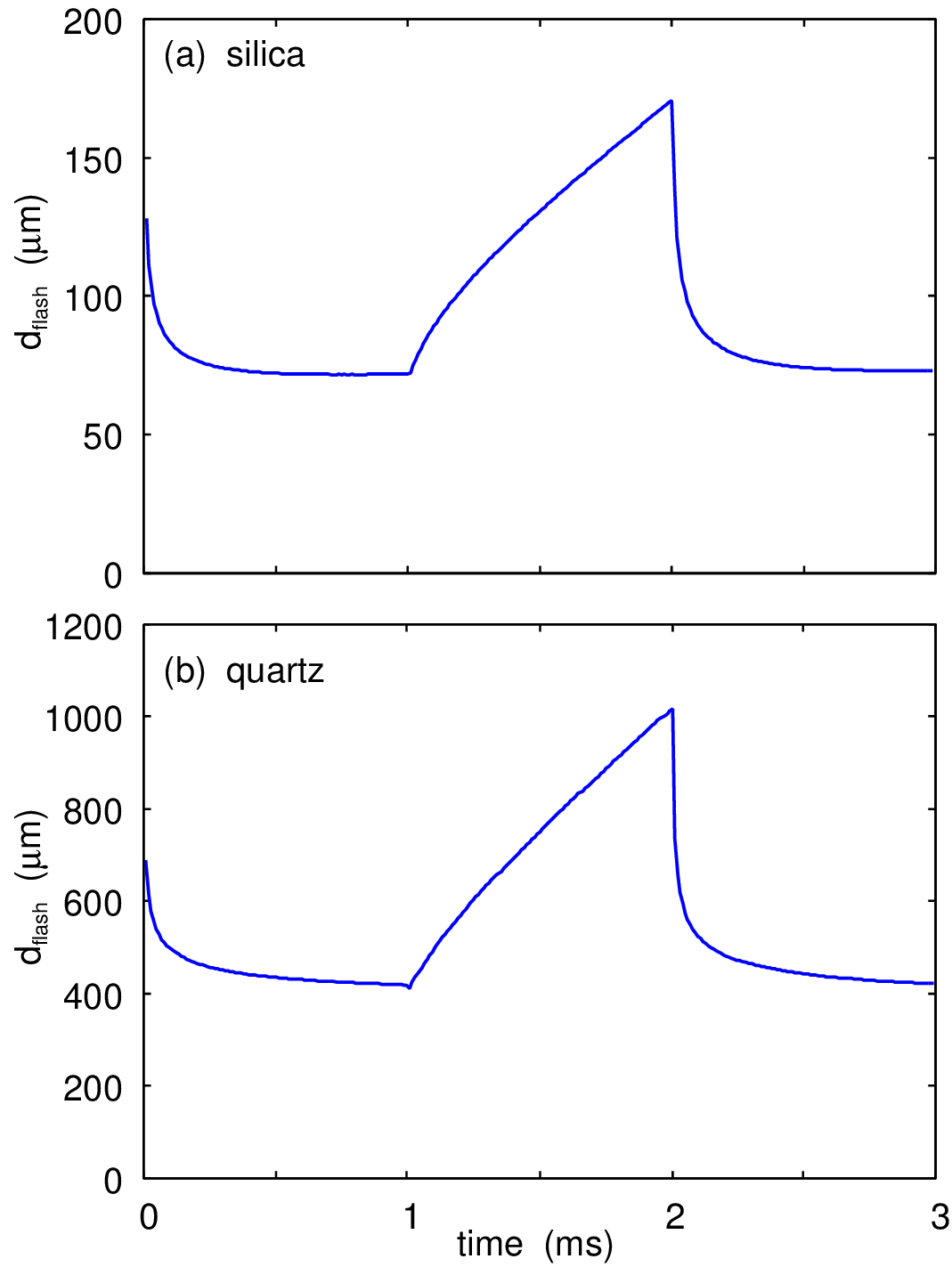}
\caption{\label{1time.2dflash.silica.eps}
The slope length $d_{\rm flash}(t)$  as a function of time for (a) silica
and (b) quartz.
}
\end{figure}

Fig. \ref{1time.2dflash.silica.eps}
shows the slope length $d_{\rm flash}(t)$  as a function of time for (a) silica
and (b) quartz. The slope length is much larger for silica than for quartz. Since the 
effective diameter of the hot spots are similar during sliding the much larger
$d_{\rm flash}(t)$ imply that the temperature distribution within the hot spots is more
isotropic for quartz than for silica. We attribute this to the higher thermal conductivity 
of quartz which makes the temperature at the front side and exit side of the contact regions 
more similar for quartz as compared to silica. During the stop time period $d_{\rm flash}(t)$ 
increase for both quartz and silica which again is due to the heat diffusion which makes the temperature profiles
more isotropic with increasing time.

\vskip 0.2cm
{\bf 4 Discussion}

The theory presented above and in Ref. \cite{MP} assumes that the frictional energy is produced in a molecular thin
(nanometer) layer at the solid surface or interface. This is the case in most applications
but not for viscoelastic solids like rubber, where an important contribution to the friction arises
from bulk viscoelastic deformations. Thus when a rubber block slides on a rigid
substrate the rubber in the substrate asperity contact regions undergoes time-dependent deformations,
which extend into the rubber a similar distance as it extends laterally. The influence of the flash
temperature on rubber friction was studied in Ref. \cite{rub1,rub2,rub3} but effectively using a one-length scale
approach where the frictional energy produced by the microasperities in the macroasperity
contact regions was smeared out within the macroasperity contact region. An extension of this theory
to include the multiscale roughness using a similar approach as in this study and in
Ref. \cite{MP} will be published elsewhere\cite{ToBe}.

\vskip 0.2cm
{\bf 5 Summary and conclusion}

I have generalized the multiscale flash temperature theory developed in Ref. \cite{MP} to non-steady (accelerated)
motion, and presented numerical results which illustrate the effect of non-steady sliding on the 
flash temperature. The theory can be applied to surfaces with roughness over arbitrary many decades in length scale
and require as input only the surface roughness power spectra and the thermal and 
the elastic (or elastoplastic) properties of the solids.

\end{document}